\documentclass[twocolumn]{article}
\usepackage{epsfig}
\title{\bf When it Comes to Spintronics, There May be Some Room in the Middle }
\author{ Supriyo Bandyopadhyay \\
Department of Electrical Engineering and Department of Physics \\
Virginia Commonwealth University, Richmond, VA 23284, USA \\
E-mail: sbandy@vcu.edu}
\date{}
\begin{document}
\maketitle
\thispagestyle{empty}
\begin{center}
{\bf ABSTRACT}
\end{center}
\medskip
Spintronic devices that utilize the spin degree of freedom of a charge carrier 
to store, process or transmit information,  may be better performers than their 
traditional electronic counterparts if special properties of ``spin'' are 
exploited in the design.  I review here the example of {\it single spin logic} 
which follows this principle. It can reduce power dissipation by several orders 
of magnitude and is the progenitor of today's spin based quantum logic gates. 
These devices are far easier to realize than ``quantum computers'', while  much
more likely to exhibit the advantages touted for spintronics than mere 
spin-based analogs of conventional  transistors.

\medskip

\begin{center}
{\bf INTRODUCTION}
\end{center}

\medskip

Authors of spintronic papers frequently introduce their work with the clich\`e 
that spin based devices will be superior to their electronic counterparts in  
power and speed. This dogmatic belief is now being increasingly questioned 
\cite{dyakonov,bandy1,bandy2}. Spintronic analogs of classical field effect 
\cite{datta}, or bipolar junction transistors \cite{dassarma,flatte} are not 
likely to offer much advantage unless revolutionary advancements are made in 
spintronic materials. Recently, we showed that  spin-based transistors  
actually lose to 
traditional  transistors  in almost every respect \cite{bandy1,bandy2}. This 
then begs the question as to where, if anywhere at all, spin devices will have 
an advantage.
To answer this question, one needs to first establish {\it why} spin based 
devices could  have an advantage in the first place. 

In the context of classical switching devices, ``spin'' has a fundamental 
advantage. Most charge based digital devices  are switched  by moving charges 
in space. In the case of a normal field effect transistor, charge is moved into 
the channel to turn the
device ON, and then out of the channel to turn the device OFF. A finite amount 
of work is done to physically move charges and this is ultimately dissipated as 
heat.  In the case of bipolar junction transistors,  the dissipation is 
associated with motion of charges into and out of the base of the transistor. 
Suffice it to say then that in order to reduce dissipation, {\it one must 
eliminate physical motion of charges}.

Imagine now a scenario where the spin orientation of a single electron becomes 
a bistable quantity. This is achieved by  placing an
electron in a quantum dot and applying a weak magnetic field so that the 
Hamiltonian describing this 
electron becomes
\begin{equation}
H = \left ( {\vec p} - e {\vec A} \right )^2/2m^* - (g/2) \mu_B {\vec B} \cdot 
{\vec \sigma}
\end{equation}
where ${\vec B}$ is the magnetic field and ${\vec A}$ is the associated vector 
potential. If the magnetic field is directed along the z-direction (${\vec B} = 
B \hat{z}$), then diagonalization of the above Hamiltonian immediately yields 
two mutually
orthogonal eigenspinors [1, 0] and [0, 1] which are states with their spin
quantization axes directed parallel and anti-parallel to the z-directed 
magnetic field.
Thus, the spin quantization axis (or spin polarization) becomes a {\it binary} 
variable -- ``down'' (parallel to the field), or ``up'' (anti-parallel to the 
field). These two  states can encode logic bits 0 and 1, respectively. 

We could switch from logic bit 0 to 1, or vice versa, by simply flipping spin
{\it without physically moving charges}. Therefore,
we expect the switching to be accomplished with minimal energy dissipation.
The dissipation will be of the order of the energy separation between the two 
non-degenerate states (=$|g \mu_B B|$). If we assume realistic parameters such 
as $|g|$ = 20 $\footnote{For bulk InAs, $|g|$ = 15, but the g-factor in quantum 
dots can be engineered somewhat by changing the shape and size \cite{schrier}}$ 
and $B$ = 1 Tesla, then the energy dissipation is about 1 meV.
If the switching delay is about 1 $\mu$sec (spin flip processes in quantum dots 
are rather inefficient), then the power dissipated in a bit flip is 
1.6$\times$10$^{-16}$ Watts. With current technology, we can expect to produce 
about 10$^{11}$ dots/cm$^2$, so that the worst case energy dissipation per unit 
area -- the quantity device engineers are concerned with -- is a paltry 16 
$\mu$W/cm$^2$
with a device density of 10$^{11}$/cm$^2$. This is several orders of magnitude 
better than the projection of the International Technology Roadmap for 
Semiconductors for the year 2010 \cite{itrs}.  The energy advantage accrues 
from exploiting a basic feature of spin: we can flip spin without moving 
charges. This saves an enormous amount of energy.

{\bf Unwanted bit flips and error rates:}
How stable are single electron spins in quantum dots? Unwanted spin flip 
processes are strongly suppressed in these systems \cite{khaetskii, khaetski1}.
Calculations for typical GaAs quantum dots have shown that the spin flip  time 
is several hundreds of $\mu$s. 
This quantity was measured in a single electron GaAs
quantum dot  and was found to exceed 50 $\mu$s in a magnetic field of 7.5 Tesla 
and 14 Tesla, with no indication of a magnetic field dependence \cite{hanson}. 
Thus, we expect unwanted bit flips to occur at low enough rates that 
conventional error correction schemes can easily handle them.

\begin{figure}[h]
\epsfig{file=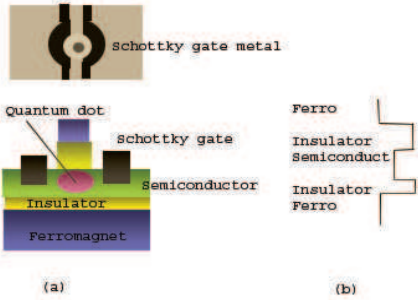,height=5.1cm,width=7.5cm,clip=,bbllx=62pt,
bblly=67pt,bburx=580pt,bbury=454pt}
\caption{(a) The structure for a gated quantum dot to host a single spin. The 
top figure  shows the top view and the bottom figure the cross-section. (b). 
The conduction band energy diagram in the direction perpendicular to the 
heterointerfaces (at equilibrium). \label{dot}} 

\end{figure}

{\bf Reading and writing of spin:} In order to  controllably orient and detect 
spin polarization of a single electron in a quantum dot (i.e. read and write a 
bit), we have to design and fabricate the structure appropriately. The quantum 
dots are delineated electrostatically in a penta-layered structure consisting 
of a ferromagnet-insulator-semiconductor-insulator-ferromagnet combination as 
shown in Fig. \ref{dot}(a). A wrap-around split Schottky gate delineates the 
boundary of the quantum dot in the semiconducting layer. This gate also allows 
the ``write'' operation. The ferromagnetic layers too serve multiple purposes. 
First, they apply a weak magnetic field on the electron in the semiconductor 
dot and thus define the spin quantization axis, while making spin polarization 
a bistable quantity. Second, they aid in performing the ``reading'' and 
``writing'' operations. 

The materials are so chosen that the conduction band energy diagram at 
equilibrium (in the direction normal to the heterointerfaces) is as shown in 
Fig. \ref{dot}(b). ``Writing'' spin, or aligning its polarization either 
parallel or anti-parallel to a magnetic field, is achieved as follows. At 
first, the lowest subband edge in the semiconductor is above the Fermi level in 
the metallic ferromagnets, so that the quantum dot is completely depleted of 
any electron. In order to write a spin, a positive potential is applied to the 
wrap-around gate which decreases the confinement and effectively makes the 
semiconductor dot larger, thereby pulling the subband edge below the Fermi 
level in the ferromagnet. A single electron now tunnels into the semiconductor 
from a ferromagnet. This electron's spin is that of the majority carriers in 
the ferromagnet. If the ferromagnet is a half metal with 100\% spin 
polarization, and is magnetized in a known direction, then we know what the 
spin polarization of this electron is.  Hence,
we have ``written'' a bit.

What if this polarization was not the desired polarization and we had intended 
to write the opposite polarization? Then, we have to flip this bit. For this 
purpose, we
apply a differential potential between the two Schottky gates to induce a 
Rashba interaction in the dot. The total spin-splitting energy in the 
semiconductor 
layer is now \cite{superlattice2}
\begin{equation}
\Delta_{s} = 2 \sqrt{ \left ( {{g \mu_B B}\over{2}}  
\right )^2 
+  {\cal E}_y^2  { {16 e^2\hbar^4}\over{ m^{*4}c^4 W_x W'_x} }
f ( W_x , W'_x ) }
\label{rashba}
\end{equation}
where the function $ f ( W_x , W'_x ) $ is given by
\begin{eqnarray}
f ( W_x , W'_x ) & = & cos^2 \left ({{\pi W_x}\over{2 W'_x}} \right ) 
cos^2 \left ({{\pi W'_x}\over{2 W_x}} \right ) F ( W_x , W'_x ) \nonumber \\
F ( W_x , W'_x ) & = & {{1 }\over{\left ( (W'_x/W_x)^2 -1 \right )^2 \left ( 
(W_x/W'_x)^2 -1 \right 
)^2}} ~,
\end{eqnarray}
where ${\cal E}_y$ is the electric field due to the differential potential 
between the Schottky gates, $W_x$ is the spatial width of the wavefunction in 
the lower spin state, and $W'_x$ is the spatial width in the upper spin state 
(they are different because the potential barriers confining the electron are 
of 
finite height). 

By adjusting  ${\cal E}_y$, we can make the total spin splitting energy in the 
chosen dot resonant with a global ac magnetic field with frequency $\omega$
($\Delta_{s}$ = $\hbar \omega$). We hold this resonance for a 
time duration $\tau$ such that 
$(2/h) \mu_B B_{ac} \tau$ = 1,
where $B_{ac}$ is the amplitude of the ac magnetic field. This flips the spin 
and allows us to write the desired bit.

The Rashba interaction that occurs during the gating operation can cause spin 
relaxation in a quantum dot. Therefore, the writing operation must be completed 
quickly. In other words, we need $\tau$ to be much less than 1 $\mu$s. If 
$\tau$ = 0.1 $\mu$s, then $B_{ac}$ = 3.6 gauss, which is very reasonable.

How much energy is dissipated during the gating operation? It is of the order 
of $CV^2$, where $C$ is the dot capacitance and $V$ is the voltage applied 
to the dot. If we assume that ${\cal E}_y$ = 10$^5$ V/cm and the dot dimension 
is 10 nm, then $V$ = 100 mV. If $C$ = 1 aF, then the energy is 10$^{-20}$ 
Joules. If the gating operation is completed in 0.1 $\mu$s, then the power 
dissipated is 10$^{-13}$ Watts, which is still very small.

``Reading'' a spin, or ascertaining its polarization, is more difficult than 
writing spin. Single spin reading has been demonstrated with magnetic resonance 
force microscopy \cite{rugar}, but this is difficult and slow. For electrical 
detection, we can use the technique of ref.  \cite{prb} (which is by no means 
unique and variations exist).

\medskip
\begin{center}
{\bf SINGLE SPIN LOGIC CIRCUITS}
\end{center}
\medskip

The idea of using an entity like spin to encode information and perform 
computation is probably not very new. However, to my knowledge, no concrete 
scheme for actually doing this existed before 1994. In 1994, two of my 
colleagues and I presented the first design for a  universal NAND gate using 
three interacting single electron spins in three quantum dots 
\cite{nanotechnology}. The NAND gate is universal; therefore, any combinational 
or sequential circuit can be realized with it. Below, I repeat the basic idea.

\begin{figure}[h]
\epsfig{file=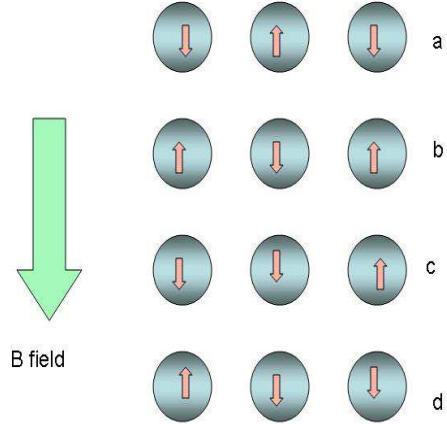,height=7.5cm,width=7.5cm,clip=,bbllx=74pt,
bblly=76pt,bburx=490pt,bbury=424pt}
\caption{Spin configurations in a 3-dot system with nearest neighbor exchange 
coupling. The peripheral spins are the two inputs and the central spin is the 
output. The four configurations correspond to the four entries in the truth 
table of a NAND gate.   \label{NAND}} 
\end{figure}

Consider a linear chain of three electrons in three quantum dots. Each electron 
is placed in a magnetic field because of the ferromagnetic contacts around it. 
As mentioned before, the spin polarization is a bistable quantity aligned 
either parallel or anti-parallel to the field. We assume that only nearest 
neighbor electrons interact via exchange since their wavefunctions overlap. We 
will also assume that the exchange energy (defined as the energy difference 
between the triplet and singlet states of two neighboring interacting 
electrons) is larger than the Zeeman splitting energy. In this
case, the ground state of the system has {\it anti-ferromagnetic ordering} 
where nearest neighbors have opposite spin. In fact, the ground state of the 
array looks like as in Fig. \ref{NAND}(a). 

Let us now regard the two peripheral spins as the two ``inputs'' to the logic 
gate and the central spin as the ``output''. Since the downspin state 
($\downarrow$) represents logic 0, and the upspin state ($\uparrow$) is logic 
1, we find that when the two inputs are 0, the output is automatically 1 
because of the anti-ferromagnetic ordering. That is encouraging since it is one 
of the four entries in the truth table of a NAND gate.

We now have to realize the other three entries in the truth table. If  we 
change the two inputs to 1 from 0, then this takes the system to an excited 
state. We then let the system relax. In order to maintain the 
anti-ferromagnetic ordering, either the central spin will flip down, or else 
the two peripheral spins will flip down to their original state after the 
writing operation is complete. The former requires a {\it single} spin flip, 
while the latter requires {\it two} spin flips. However, the former process 
will take the system to a local metastable state, while the latter takes it to 
the global ground state. 

Whether the metastable state is reached, or the global ground state is reached
depends on the energy landscape. If the metastable state is reached, then we
will achieve the configuration shown in Fig. \ref{NAND}(b). This is the desired 
configuration because when the two inputs are 1, the output is 0. This is yet 
another entry in the truth table of a NAND gate. 

The metastable state, left to itself, must ultimately decay to the global 
ground state. But if the energy barrier between the two states is large enough, 
this process may take very long. If it is much longer than the inverse of the 
input data rate, we can ignore it.

How do we guarantee that the metastable state is reached and not the global 
ground state? We can never do it with 100\% certainty, but we can do it with a 
high degree of certainty if we make the two-spin-flip process much more 
inefficient than the single-spin-flip process. Since selection rules are rather 
stringent in quantum dots, this is a very likely scenario.

Finally, what happens if one input is 1 and the other 0? This situation 
seemingly causes a {\it tie}, but the weak magnetic field present in every dot 
causes a Zeeman interaction which resolves this situation in favor of the 
central dot having a down-spin configuration. This situation is shown in Figs.  
\ref{NAND}(c) and \ref{NAND}(d). Note that these conform to the other two 
entries in the 
truth table of a NAND gate.

Finally, we have realized the entire truth table:

\medskip
\begin{center}
\begin{tabular}{|c|c|c|}
\hline
Input 1 & Input 2 & Output \\
\hline
0 & 0 & 1 \\
1 & 1 & 0 \\
0 & 1 & 0 \\
1 & 0 & 0 \\
\hline
\end{tabular}
\end{center}
\medskip

In 1995, Molotkov and Nazin verified the entire truth table using a fully 
quantum mechanical exact many body calculation \cite{molotkov}.  Further 
calculations of this system have been carried out by Bychkov and co-workers 
\cite{bychkov}.

{\bf The all-important issue of unidirectionality:}
There is, however, a serious problem with this type of logic gates which may 
not be apparent to the untrained. There is no
{\it isolation} between the input and output of this type of logic gate since 
the exhange interaction is bidirectional. It does not distinguish between which 
spin is the input bit and which is the output. This makes it impossible for 
logic signal to flow {\it unidirectionally} from an input stage to an output 
stage and not the other way around.  We have discussed this
 at length in various publications 
\cite{nanotechnology,jjap,egypt,superlattice3} since it is vital. In 1994, when 
we first proposed these logic gates \cite{nanotechnology}, we proposed to 
enforce unidirectionality by progressively increasing the distance between 
quantum dots, so that there is  
{\it spatial} symmetry breaking.  In 1996, we revised this idea and explored  
imposing unidirectional flow of signal in {\it time}, rather than in space, by 
using clocking \cite{jjap}. This is actually commonplace in bucket brigade 
shift registers realized with charge coupled devices, where a push clock and a 
drop clock are used to steer a charge packet unidirectionally from one device 
to the next. At that time, 
we realized that a single phase clock cannot impose the required 
unidirectionality. Later, we found that a 3-phase clock is required to do this 
job \cite{superlattice3}.  The reader is  
referred to \cite{nanotechnology,jjap,egypt,superlattice3,anant} for a 
discussion of this important topic.

The clocking circuit however introduces additional dissipation. The energy 
dissipated in a clock cycle, in a single clocking line, is about $C'V_{c}^2$, 
where $C'$ is the capacitance of the 
clocking line and $V_c$ is the voltage swing in the line. If we assume $C'$ = 1 
fF, $V_c$ = 100 mV, then the energy dissipated is 10$^{-17}$ Joules. The clock 
frequency is about 1 MHz (of the same order as the spin flip rate), so that the 
power dissipation in a clock line is of the order of 10$^{-11}$ Watts. This is 
still a small quantity.

\medskip
\begin{center}
{\bf ADIABATIC/REVERSIBLE GATES}
\end{center}
\medskip

So far, we have discussed a spintronic logic family that dissipates very little 
energy. But can we design logic gates that dissipate no energy at all? It is 
well known that such gates must be {\it logically reversible} \cite{landauer}, 
i.e. we should be able to infer the input unambiguously from the output state. 
In 1996, we presented such a gate which is a {\it quantum adiabatic inverter} 
\cite{superlattice}. Just two exchange coupled spins in a weak magnetic field 
make an inverter, which is logically reversible since the input can always be 
inferred from the output (because they are simply logic complements of each 
other). Therefore, the inverter could be switched adiabatically without 
dissipating any energy at all.

In ref. \cite{superlattice}, we presented the adiabatic inverter. Its dynamics 
is governed by the time-dependent Schr\"odinger equation and the gate is both 
logically and physically reversible, dissipating no energy at all.
We showed that once a fresh input is applied to the inverter, the output 
switches to the correct state (the logic complement of the input) in a time 
$t_d$ given by
\begin{equation}
t_d = h/(4\sqrt{h_A^2 + 4 J^2}) ~,
\end{equation}
where $h$ is the Planck constant, $h_A$ is the energy applied to switch the 
input (or align the spin in the input dot to the desired orientation) and $J$ 
is the exchange splitting energy, which is the energy difference between the 
triplet and singlet states of the two electrons. Note that neither the energy 
$h_A$, nor $J$ is dissipated in the switching process. We could switch the 
device arbitrarily fast by applying arbitrarily large $h_A$, but we also showed 
in ref. \cite{superlattice}, that in order for the inverter to switch perfectly 
and behave correctly, we need
$h_A$ = $2J$,
which makes $t_d$ = $h/(8\sqrt{2}J)$. Even if $J$ = 1 meV, $t_d$ is fractions 
of a picosecond, which is reasonably fast. Since the switching delay is less 
than 1 ps and the spin coherence time in a quantum dot can easily exceed 100 
$\mu$s, this gate is suitable for fault tolerant quantum computing. Later, a 
similar idea was discussed by Openov and Bychkov \cite{openov}.

Double quantum dot systems are routinely fabricated and their spin states have 
been both measured and controlled \cite{tarucha}. Therefore, adiabatic 
inverters currently exist.

\paragraph{Toffoli-Fredkin gate:} The adiabatic inverter however is not a 
universal adiabatic gate. The universal adiabatic gate is the Toffoli Fredkin 
gate which has three inputs $A,B,C$ and three outputs $A',B',C'$. The 
input-output relationships are $A'$ = $A$, $B'$ = $B$ and 
$C'$ = $C \bigoplus A \bullet B$, where $\bigoplus$ is the EXCLUSIVE-OR 
operation and $\bullet$ is the AND operation. That means $C$ toggles iff $A$ 
and $B$ are both logic 1; otherwise, nothing happens. We can realize this gate 
with three spins in a  magnetic field with nearest neighbor exchange 
interaction 
(exactly the same configuration as the NAND gate described earlier). However, 
there is only one difference with the NAND gate. This time, we will make the 
Zeeman interaction (due to the magnetic field)
stronger than the exchange interaction, i.e. $g \mu_B B > J$.

\begin{figure}[h]
\epsfig{file=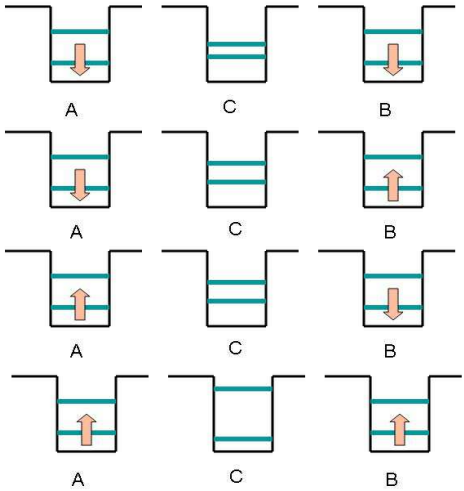,height=7.5cm,width=7.5cm,clip=,bbllx=64pt,
bblly=29pt,bburx=478pt,bbury=460pt}
\caption{The realization of a spintronic Toffoli-Fredkin gate. The spin 
splitting energies in the central dot are shown as a function of the spin 
orientations in the two peripheral dots. The two peripheral dots are the two 
control bits $A$ and $B$ and the central spin is the target bit $C$. 
\label{toffoli} }
\end{figure}

The magnetic field induces a Zeeman splitting which makes the downspin 
state lower in energy than the upspin state in each dot. The exchange 
interaction tries to make spins in neighboring dots anti-parallel. The two 
peripheral spins are the control bits $A$ and $B$ and the central spin is the 
bit $C$.

If the spins in $A$ and $B$ are ``down'' ($\downarrow$), then the exchange 
energy 
will tend to keep the spin in $C$ ``up'' ($\uparrow$), while the Zeeman 
interaction still maintains the downspin state (the state parallel to the 
magnetic field) as the lower energy state. In this case, the exchange 
interaction {\it subtracts} from the Zeeman splitting in the central dot and 
makes the total spin splitting energy in this dot less than the bare Zeeman 
splitting. If the spin in dot $A$ is ``up'' and that in dot $B$ ``down'', then 
the exchange interaction effects due to $A$ and $B$  on the spin in dot $C$ 
tend to {\it cancel} and the spin splitting in dot $C$ is more or less the 
Zeeman energy. On the other hand, if the spins in $A$ and $B$ are both ``up'', 
then the exchange interactions due to them {\it add} to the Zeeman splitting in 
dot $C$, making the total spin splitting energy in $C$ larger than the bare 
Zeeman splitting. In essence, the total spin splitting energy in $C$ is {\it 
larger} when both $A$ and $B$ are in logic 1 state, than otherwise. If we 
denote the spin splitting energy in dot $C$ as $\Delta^C_{\alpha \beta}$ where 
$\alpha$ and $\beta$ are the logic states in dots $A$ and $B$, then the 
following inequality holds:
\begin{equation}
\Delta^C_{11} > \Delta^C_{10} = \Delta^C_{01} > \Delta^C_{00}
\end{equation}
These situations are shown in the energy diagrams in Fig. \ref{toffoli}.

To implement the dynamics of the gate,  the whole system is pulsed with a 
global ac magnetic field of amplitude $B_{ac}$ whose frequency 
is resonant with $\Delta^C_{11}$. The pulse duration is $h/(2 \mu_B B_{ac})$. 
Therefore, $C$ will toggle iff $A$ and $B$ are both in logic state 1. 
Otherwise, nothing will happen. This is the standard technique for realizing a 
Toffoli-Fredkin gate (see ref. \cite{lloyd} for a similar idea).

\medskip
\begin{center}
{\bf QUANTUM COMPUTING WITH SPINS}
\end{center}
\medskip

The idea of using spins in quantum dots to encode qubits was already latent in 
 ref. \cite{superlattice}. Subsequently, a series of proposals followed, 
articulating the use of spins in different systems to encode qubits and 
implement universal quantum gates \cite{privman, loss, kane, bandy_prb}. The 
advantage of using spin, as opposed to charge, as hosts for qubits is that spin 
coherence is more robust than charge coherence. 
This field is currently an arena of active research and some progress is 
expected in near future, although practical quantum computers with the 
capability to entangle and maneuver several qubits may be decades away.

\medskip
\begin{center}
{\bf CONCLUSION}
\end{center}
\medskip

In this short review, I have presented a genre of low power devices where 
switching is accomplished by flipping spin rather than physically moving 
charges. The ultimate embodiment of low power circuits are quantum computers 
which do not dissipate any power at all. These are, of course, at the top of 
the hierarchy, but also the most challenging and difficult to realize. At the 
bottom of the hierarchy are the spin based analogs of conventional field effect 
and bipolar transistors, which do not promise significant advancements at this 
time. In the middle is the single spin (classical) logic family which is less 
difficult to realize than quantum computers and more promising than mere 
spintronic analogs of transistors.
There may be plenty of room in the middle for exciting research.

\medskip
{\bf Acknowledgement}
\medskip

The author's work in spintronics is supported by the US Air Force Office of 
Scientific Research under grant FA9550-04-1-0261.

\end{document}